# Large phonon band-gap in SrTiO$_3$ and the vibrational signatures of ferroelectricity in ATiO$_3$ perovskite: First principles lattice dynamics and inelastic neutron scattering of PbTiO$_3$, BaTiO$_3$ and SrTiO$_3$.

Narayani Choudhury[1], Eric J. Walter[2], Alexander I. Kolesnikov[3] and Chun-Keung Loong[3]

[1]Solid State Physics Division, Bhabha Atomic Research Centre, Trombay, Mumbai 400085, India
[2]College of William and Mary, Williamsburg, Virginia 23187, USA
[3]Intense Pulsed Neutron Source Division, Argonne National Laboratory, Argonne, Illinois 60439, USA

**Abstract**

We report first principles density functional perturbation theory calculations and inelastic neutron scattering measurements of the phonon density of states, dispersion relations and electromechanical response of PbTiO$_3$, BaTiO$_3$ and SrTiO$_3$. The phonon density-of-states of the quantum paraelectric SrTiO$_3$ is found to be fundamentally distinct from that of ferroelectric PbTiO$_3$ and BaTiO$_3$ with a large 70-90 meV phonon band-gap. The phonon dispersion and electromechanical response of PbTiO$_3$ reveal giant anisotropies. The interplay of covalent bonding and ferroelectricity, strongly modulates the electromechanical response and give rise to spectacular signatures in the phonon spectra. The computed charge densities have been used to study the bonding in these perovskites. Distinct bonding characteristics in the ferroelectric and paraelectric phases give rise to spectacular vibrational signatures. While a large phonon band-gap in ATiO$_3$ perovskites seems a characteristic of quantum paraelectrics, anisotropy of the phonon spectra correlates well with ferroelectric strength. These correlations between the phonon spectra and ferroelectricity, can guide future efforts at custom designing still more effective piezoelectrics for applications. These results suggest that vibrational spectroscopy can help design novel materials.

**I. Introduction**

Ferroelectric materials interconvert electrical and mechanical energies and find key technological applications[1] as piezoelectric transducers and actuators used in ultrasonic devices, medical imaging and telecommunications. Classic perovskites like PbTiO$_3$, BaTiO$_3$ and SrTiO$_3$ present unusual properties and puzzles[1-23] and their first principles calculations[2-12] and experiments[13-23] have helped elucidate a wide range of fundamental issues like the electronic origins of ferroelectricity[3], soft phonon modes and structural phase transitions[1,13], giant LO-TO (longitudinal optic-transverse optic) splittings[7] and vibrational anomalies[4-9], origin of ultrahigh electromechanical response[10], *etc*. SrTiO$_3$ is an incipient ferroelectric with a very large static dielectric response, which exhibits remarkable phonon anomalies[5,18] and electrostrictive response[19]. The ferroelectric phase in SrTiO$_3$ is suppressed even as T$\rightarrow$ 0 K by zero-point fluctuations leading to quantum paraelectricity[15].

Both PbTiO$_3$ and BaTiO$_3$ have a simple cubic high temperature paraelectric phase which transforms to a ferroelectric tetragonal phase around 763 K and 403 K, respectively[1]. Tetragonal PbTiO$_3$ is a large strain material ($c/a=1.06$) which is ferroelectric even at high temperatures, and exhibits a single cubic to tetragonal transition. BaTiO$_3$ on the other hand has a much smaller strain (1.01) and exhibits successive phase transitions from cubic to tetragonal, orthorhombic and rhombohedral structures with decreasing temperature. SrTiO$_3$ undergoes a transition from the cubic ($Pm\bar{3}m$) to a tetragonal ($I4/mcm$) antiferrodistortive phase at 105 K; however this transition has a non-polar character and does not affect its dielectric properties. Electronic structure calculations[3] reveal that intrinsic differences in the bonding in tetragonal BaTiO$_3$ and PbTiO$_3$, give rise to their vastly different phase diagram and ferroelectric behavior.

In this work, we report *ab initio* lattice dynamics calculations and inelastic neutron scattering studies of the complete phonon dispersion relations, density of states and electromechanical response of three classic perovskites: ferroelectric PbTiO$_3$ and BaTiO$_3$ and the quantum paraelectric SrTiO$_3$. The experimental and theoretical determination of the phonon density of states and dispersion relations gives access to valuable quantitative information concerning elasticity, piezoelectric and dielectric behavior, thermodynamic properties and the dynamics of soft-mode driven phase instabilities. The fundamental interest, distinct ferroelectric behavior and important applications make PbTiO$_3$, BaTiO$_3$ and SrTiO$_3$ highly suitable for these studies. Our goals are, (i) to understand the vibrational signatures of ferroelectricity, (ii) to study the interplay between the structure, bonding, dynamics and electromechanical response, and, (iii) to identify the factors that govern enhanced piezoelectric response, required for the design of new materials. The neutron measurements provide a critical test for the theory. The integration of first principles calculations and vibrational spectroscopic experiments provide important insights on the correlations between vibrational spectra and ferroelectricity, and illustrate how vibrational spectroscopic techniques can lead to the design of novel materials.

Several workers[20] have reported inelastic neutron scattering measurements of the phonon dispersion relations and have studied the temperature variations of the soft phonon mode across the ferroelectric to paraelectric transition in these perovskites. The phonon dispersion relations data are however incomplete especially in the lower symmetry ferroelectric phases. Even in the high symmetry paraelectric cubic phase, the dispersion relations of only the acoustic and low frequency optic phonon branches have been measured[20]. Thus far, first principles calculations of the complete phonon dispersion relations are reported only for the cubic phases[2,6] of PbTiO$_3$, BaTiO$_3$ and SrTiO$_3$. A thorough understanding of the phonon dispersion relations in the ferroelectric phases is essential for a microscopic understanding of their electromechanical properties and for resolving the controversies about the origin of diffuse scattering[23] in perovskite ferroelectrics and have formed the focus of this work. In spite of a flurry of reported research on these perovskites[1-23], studies of the complete phonon density of states and systematic examinations of the vibrational signatures of ferroelectricity, etc. were not earlier studied. Systematic examination of these various issues which form the fundamental link between the microscopic physics and macroscopic material behavior is the key to defining new experimental protocols for the screening and design of novel functional materials. Classic perovskites like PbTiO$_3$, BaTiO$_3$ and SrTiO$_3$ which have served as a fertile ground for the discovery of new physical phenomena and devices[1-23] are ideal model systems for such studies.

## II. Techniques
### A. Theoretical studies

We have undertaken first principles calculations of ferroelectric tetragonal PbTiO$_3$ and rhombohedral BaTiO$_3$ with space groups *P4mm* and *R3m* and the paraelectric cubic (*Pm$\bar{3}$m*) and antiferrodistortive tetragonal phases (*I4/mcm*) of SrTiO$_3$ (ST). Density functional theory (DFT) permits calculation of the total energy of solids without any parameterization to experimental data. The phonon spectra, elastic, piezoelectric and dielectric tensors are related to the second derivates of the total energy with respect to variables like atomic displacements, macroscopic strain, and electric field. All these material properties can be efficiently computed using density functional perturbation theory (DFPT)[24-27]. The dielectric susceptibility and elastic constants involve second derivatives of total energy with respect to electric field and strain, respectively, while the piezoelectric tensor is the mixed second derivative of total energy with respect to strain and electric field. DFPT linear response studies have intrinsic advantages over the frozen phonon technique, as they do not require large supercells for the studies of phonons at a general wavevector. Furthermore, the DFPT approach includes explicit treatment of the long ranged Coulomb interactions, required to obtain accurate values of the LO-TO splittings.

We carried out DFT and DFPT linear response calculations using plane wave basis sets and the code ABINIT[28] using the local density approximation (LDA). These calculations used norm-conserving pseudopotentials[29-32] generated using the code OPIUM. The pseudopotential results were rigorously tested[31] against the full-potential linearized augmented plane-wave calculations and included semicore states of Pb $5d^{10}$, Ti $3s^2 3p^6 4d^2$, and O $2s^2$ as valence states. The Brillouin zone integrations were performed with a 6x6x6 **k**-point mesh using a plane-wave energy cut off of 120 Rydbergs. We first carried out full structural relaxations for these perovskites. Linear response DFPT studies[24-26] using a 6x6x6 **k**-point grid with atomic displacements, strain and electric field perturbations were used to compute the zone center phonon frequencies, elastic, piezoelectric and dielectric properties. The spontaneous polarization was computed using the Berry's phase approach[27] using a 6x6x20 **k**-point grid, with a dense mesh along the direction of polarization. All calculations are at zero temperature.

To study the phonons at a general wavevector, response functions were calculated on a 4x4x4 grid of **q**-points in the Brillouin zone (BZ), including the zone center Γ point. The phonon frequencies at a general wavevector **q** were obtained by interpolating the dynamical matrices calculated on this grid. The phonon density of states involves an integrated average over the phonon modes in the entire Brillouin zone and is given by,

$$g(\omega) = A \int_{BZ} \sum_j \delta(\omega - \omega_j(\mathbf{q})) d\mathbf{q}$$

where A is a normalization constant such that $\int g(\omega) d\omega = 1$ and $\omega_j(\mathbf{q})$ is the frequency of the $j^{th}$ phonon mode. We used a 12x12x12 **q**-point mesh in the irreducible Brillouin zone to calculate the phonon density of states.

### B. Experimental studies

We have carried out inelastic neutron scattering measurements (T=6 K) of the phonon density of states of three classic perovskites: ferroelectric PbTiO$_3$ and BaTiO$_3$ and the quantum paraelectric SrTiO$_3$. The inelastic neutron scattering measurements were carried out using powder samples on the High-Resolution Medium-Energy Chopper Spectrometer (HRMECS) using the time-of-flight technique at Argonne's Intense Pulsed Neutron Source. Two incident neutron energies were used (50 and 130 meV) to obtain good resolution data in all range of energy transfers. The data were collected over a wide range of scattering angles (28° to 132°) for large coverage of momentum transfers. The high energies of neutrons from pulsed sources enable measurements of the phonon spectra over its entire energy range. Samples were cooled to 6 K using a conventional liquid-helium cryostat with the sample to minimize multiphonon contributions. The energy resolution $\Delta E$ (full width at half maximum) of the HRMECS spectrometer varies between 2–6% of the incident-neutron energy ($E_0$) over the neutron-energy-loss region. The data were corrected for background scattering by subtracting the results from empty container runs. Measurements of the elastic incoherent scattering from a vanadium standard provided the detector calibration and intensity renormalization.

The observed data were analyzed in the incoherent approximation[33,34], wherein the measured scattering function $S(Q, E)$ in the neutron energy loss experiments is related to the generalized density of states by[34]

$$g^n(E) = B \left\langle \frac{e^{2W(Q)}}{Q^2} \frac{E}{n(E,T)+1} S(Q,E) \right\rangle \approx C \sum_k \left\{ \frac{4\pi b_k^2}{m_k} \right\} g_k(E)$$

where the partial density of states $g_k(E)$ is given by $g_k(E) = D \int_{BZ} \sum_j |\xi(\mathbf{q}j, k)|^2 \delta(E - E_j(\mathbf{q})) d\mathbf{q}$ and $n(E,T) = [exp(E/k_B T) - 1]^{-1}$. B, C and D are normalization constants. $b_k$ and $m_k$ are respectively, the neutron scattering-length and mass of the $k^{th}$ atom. $E_j(\mathbf{q})$ and $\xi(\mathbf{q}j)$ respectively correspond to the energy and eigenvector of the $j^{th}$ phonon mode at wavevector **q** in the *BZ*. The symbol $\langle \rangle$ represents *Q* averaging of the quantities within.

The incoherent approximation[33, 34] is valid, when the ratio of the volume of the reciprocal space covered in the experiment to the volume of the Brillouin zone is large. In our inelastic neutron scattering experiment, this ratio was about a few thousands. The data were properly averaged over the range of scattering angles to obtain the neutron-weighted generalized phonon density of states $g^n(E)$. The first principles calculations were used to derive the generalized phonon density of states $g^n(E)$ for comparison with the experiments. The computed density of states were smeared with Guassians having full width at half maximum of 1 meV for comparison with the experiments, due to the finite resolution involved in the measurements.

## III. Results

### A. Phonon dispersion relations and long wavelength phonon frequencies

The calculated phonon dispersion relations of $PbTiO_3$ (Fig. 1) reveal giant anisotropies and span distinct spectral ranges for wavevector directions along ($\Gamma$-Z-A-R-Z)) and perpendicular ($\Gamma$-M-X-$\Gamma$) to the direction of spontaneous polarization. Very limited single crystal inelastic neutron data[20] and first principles calculations[4] of the phonon frequencies of $PbTiO_3$ were earlier reported and are in good agreement with our studies (Table I). The calculated transverse acoustic and transverse optic modes of $PbTiO_3$ (Fig. 1) propagating in the direction perpendicular to the spontaneous polarization show considerable splitting due to tetragonal anisotropy in agreement with reported inelastic neutron data[20(b)]. LDA calculations of $PbTiO_3$ underestimate the strain (calculated $c/a$=1.047, observed $c/a$=1.063) and volume (calculated V=60.4 Å$^3$). As the phonon frequencies are quite sensitive to the structural parameters, we have relaxed the structural variables of $PbTiO_3$ at the observed lattice constants, as in earlier studies[4]. The computed structural parameters and long wavelength phonon frequencies of tetragonal $PbTiO_3$ and rhombohedral $BaTiO_3$ are found to be in good agreement with reported experimental data[17(c),22(a)] and LDA calculations[4,26(b)] (Table I). The phonon dispersion relations of rhombohedral $BaTiO_3$ are plotted along cubic high symmetry directions (Fig. 1), to enable direct comparisons with the $\omega_j(\mathbf{q})$ of tetragonal $PbTiO_3$.

Symmetry analysis of the zone-center $\Gamma$-point phonon modes for the *P4mm* space group of $PbTiO_3$ yields the classification $\Gamma$: $4A_1+5E+B_1$. The E symmetry phonon modes are doubly degenerate. The $A_1$ and E phonon modes which are both Raman and infrared active, are polar modes with vibrations which are respectively parallel and perpendicular to the direction of spontaneous polarization. For rhombohedral $BaTiO_3$, the phonon modes can be classified as $\Gamma$: $4A_1+5E+A_2$. The $A_1$ and 4 E modes are polar while the $A_2$ mode and a doubly degenerate E-mode are non-polar (Table I). While most of the calculated frequencies of $PbTiO_3$ are in good agreement with experiments (Table I), the observed E-symmetry polar infrared-active phonon modes[22(a)] at 289 (TO), 505 (TO) and 723 (LO) cm$^{-1}$ at the $\Gamma$ point are underestimated, which causes the shifting of the corresponding branches/peaks in the computed phonon dispersion relations and phonon density of states of $PbTiO_3$ (Figs. 1, 2). Nevertheless, LDA calculations of $PbTiO_3$ bring out all the salient features and the calculated phonon dispersion relations and density of states are overall in good agreement with experiments.

Although experimentally $SrTiO_3$ is not ferroelectric even at low temperatures, it is very close to the ferroelectric threshold. Isotopic replacement of oxygen or partial cation substitution reduces quantum fluctuations and makes it ferroelectric[14]. Path-integral Monte-Carlo simulations which include zero-point energy contributions yield the correct ground state[8] for ST. Neglect of zero-point energy in the structural relaxation, yields ferroelectric zone center instabilities for the polar $E_u$ and $A_{2u}$ phonon modes in the tetragonal *I4/mcm* antiferrodistortive structure in agreement with earlier reports[5]. LDA calculations of cubic $SrTiO_3$ yields soft zone center and zone boundary R and M point phonon instabilities in good agreement with reported all electron linear augmented plane wave calculations[6]. LDA calculations on a coarse wavevector mesh of the antiferrodistortive phase suggest that the phonon density of states of the tetragonal antiferrodistortive and cubic phases are overall quite similar.

Stirling *et al.*[20(g)] and Cowley[20(i)] have reported single crystal inelastic neutron scattering measurements at T=90 K and T=296 K of the low energy crystal dynamics and phonon dispersion relations in $SrTiO_3$. They have fitted the observed inelastic neutron data to several lattice dynamics models with temperature dependent force constants[20(g), 20(i)]. $SrTiO_3$ exhibits dynamical critical phenomena in the vicinity of the low temperature phase transition and the temperature variations of the soft mode phonon frequencies and lineshapes have aroused considerable interest[20(j)]. Our computed *ab initio* T=0 K phonon dispersion relations of cubic $SrTiO_3$ are compared with the reported T=90 K inelastic neutron scattering data[20(g), 20(i)] (Fig. 1(c)). Our calculations of cubic $SrTiO_3$ reveal very strong dispersion of the lowest energy optic phonon modes along ($\xi\xi\xi$) and ($\xi\xi 0$) wavevector directions, in good agreement with reported inelastic neutron experiments[20(g), 20(i)] (Fig. 1(c)). While the higher frequency optic modes are in satisfactory agreement, the observed zone center, R and M point soft modes in $SrTiO_3$ [which are found to be strongly temperature dependent[20(g), 20(i)]] are in qualitative agreement with our T=0 K calculations. The computed phonon density of states of cubic and tetragonal $SrTiO_3$ neglecting effects from phonon instabilities are found to be in good agreement with our measured inelastic neutron scattering T=6 K data (Fig. 2).

### B. Calculated charge densities and bonding characteristic of $PbTiO_3$, $BaTiO_3$ and $SrTiO_3$

The computed charge densities and isosurface plots of tetragonal $PbTiO_3$ and rhombhohedral $BaTiO_3$ are given in Fig. 3. All densities were computed from the fully relaxed zero pressure LDA structures. The well connected region of charge between the Pb and O atoms indicate the strong covalent Pb-O bonding in PT. In $BaTiO_3$, the charge densities almost connect (Fig. 3) while in cubic SrTiO$_3$ they are far from connecting which reveal ionic Ba-O and Sr-O bonding, respectively. Electronic structure calculations[3] reveal that the covalent bonding between the Ti and O in PbTiO$_3$ and BaTiO$_3$ arise from the hybridization between the titanium *3d* states and the oxygen *2p* states. This covalent Ti-O bonding was found to be essential for ferroelectricity in perovskites[3]. The strong covalency of the Pb-O bonds in tetragonal PbTiO$_3$ which arises from the hybridization of the Pb *6s* state and O *2p* state has been theoretically predicted[3] as a key factor of the much larger ferroelectricity of PbTiO$_3$ as compared to BaTiO$_3$. The covalent character of the Pb-O bonds have been experimentally verified[16] and the computed charge densities are in good agreement with the observed density distributions of tetragonal PbTiO$_3$ and BaTiO$_3$ obtained from maximum entropy analysis of synchrotron data. The covalent nature of the Pb-O bond stabilizes the ferroelectric tetragonal phase in PbTiO$_3$, while the ionic nature of the Ba-O bond stabilizes the rhombohedral phase of BaTiO$_3$. These structural and bonding changes in these materials lead to important differences in their PDR (Fig. 1), PDOS (Fig. 2) as well as elastic, piezoelectric and dielectric properties (Table II).

**C. Elastic, piezoelectric and dielectric properties**

The computed elastic, piezoelectric and dielectric properties of PbTiO$_3$ at the experimental volume (Table II) are in good agreement with reported room temperature Brillouin scattering data[35]. These demonstrate the intrinsic ability of DFPT calculations in accounting for acoustic phonons and polarization accurately. Electronic structure calculations[3] reveal that the hybridization between the titanium *3d* states and the oxygen *2p* states is essential for ferroelectricity in perovskites. The strong covalent character of the Pb-O bond (Fig. 3) in PbTiO$_3$ enhances its ferroelectric strength[3] as compared to BaTiO$_3$, wherein the Ba-O bonding is ionic. These bonding changes[3] between PbTiO$_3$ and BaTiO$_3$ give rise to significant differences in their Curie temperature, phase diagram and electromechanical response (Table II). The large difference between the elastic constants $C_{33}$ and $C_{11}$ (Table II) in tetragonal PbTiO$_3$ reflects its inherent large anisotropy. This anisotropy is due to the large difference in the longitudinal acoustic phonon wave velocities, for propagation along and perpendicular to the direction of polarization (Fig. 1). This anisotropy in PbTiO$_3$ leads to interesting directional enhancements of its piezoresponse (Table II). The enhanced values for $e_{33}$ and $e_{15}$ in tetragonal PbTiO$_3$ are due to the large atomic response to the corresponding macroscopic strains combined with large anomalous values of the Born effective charge tensors.

The computed partial density of states giving the dynamical contributions from various atoms are given in Fig. 4. These reveal that the 0-20 meV low frequency vibrations which critically govern the electromechanical response are strongly influenced by the A-O bonding character. The strong covalent bonding of the Pb-O bonds (Fig. 3) contributes to the strong anisotropy of the elastic and piezoelectric response of PbTiO$_3$ (Table II). Although first principles calculations of the piezoelectric response in PbTiO$_3$ and BaTiO$_3$ have also been reported by others[11,26], in this work we have examined the intimate connections between phonon spectra and electromechanical response. The computed elastic, piezoelectric and dielectric constants (Tables I, II) are found to be quite sensitive to the relaxed structural parameters and although our results for PbTiO$_3$ are similar to earlier first principles results[11], our results for BaTiO$_3$ have some differences with reported values[26(b)] which are due to the differences in their computed crystal structures.

**D. Phonon density of states and partial density of states**

The computed generalized phonon density of states is in good agreement with the observed inelastic neutron scattering spectra (Fig. 2) and spans the spectral range from 0-120 meV. The phonon spectrum of the quantum paraelectric SrTiO$_3$ is found to be fundamentally distinct from those of ferroelectric PbTiO$_3$ and BaTiO$_3$, with a large 70-90 meV phonon band-gap. This large phonon band-gap is due to the distinct bonding in SrTiO$_3$ (Fig. 3) as compared to ferroelectric PbTiO$_3$ and BaTiO$_3$. From the viewpoint of crystal stability[9,14], the quantum paraelectricity in the threshold of ferroelectric behavior in SrTiO$_3$ originates from the critical status of this material. Such trends are conventionally studied[9,14] using the tolerance factor ( $t=(r_{Sr}+r_O)/\sqrt{2}(r_{Ti}+r_O)$). Using reported ionic radii, the tolerance was found to be $t=1.00$ for SrTiO$_3$, which means that the ion packing in SrTiO$_3$ is ideal for a perovskite-type structure[14]. The tolerance $t$ for BaTiO$_3$ and PbTiO$_3$ are similarly[9], 1.07 and 1.03. Larger ($t>1$) and smaller ($t<1$) values were found to favor ferroelectricity [e.g. BaTiO$_3$] and quantum paraelectricity [eg, see Ref. [9]: CaTiO$_3$, $t=0.97$], respectively. The critical status of SrTiO$_3$ and its distinct bonding as compared to PbTiO$_3$ and BaTiO$_3$, result in spectacular signatures in the phonon spectra.

The computed partial density of states (Fig. 4) enables microscopic interpretations of the observed data. These reveal that while the A-cations (namely, the Pb, Ba and Sr) contribute in the 0-20 meV spectral range (Fig. 4), the intermediate and high energy spectra are due to the Ti and O vibrations. The phonon spectra (neglecting instabilities) are overall similar in the paraelectric cubic and tetragonal antiferrodistortive phase of SrTiO$_3$ as the distortions for the SrO$_{12}$ and TiO$_6$ polyhedra are small in the antiferrodistortive tetragonal phase of SrTiO$_3$. On the other hand, the covalent Ti-O bonding, found essential for ferroelectricity in perovskites, causes a strong distortion of the Ti-O$_6$ octahedra and the spectra are significantly different in the ferroelectric and paraelectric phases of PbTiO$_3$ and BaTiO$_3$. For example, in tetragonal PbTiO$_3$, the calculated Ti-O bond lengths (Å) are 1.9785 (4), 1.7606 (1), 2.3908 (1), [where the values in parentheses give the multiplicity of the bonds] which deviate significantly from the typical Ti-O bond length of about 1.95 Å in cubic perovskites. This strong TiO$_6$ octahedral distortion (Table I) characteristic of ferroelectric behavior is responsible for the filling up of the characteristic 70-90 meV phonon band-gap of the quantum paraelectric SrTiO$_3$. Strong covalency of the Pb-O bonds, similarly distorts the PbO$_8$ polyhedra (2.8016(4) Å; 2.5243 (4) Å) in tetragonal PbTiO$_3$, which lead to the following: (i) The vibrations of the Pb atoms in PbTiO$_3$ (around 7.5 meV) are considerably softer than in the vibrations of Ba (around 12.5 meV) in BaTiO$_3$ and Sr (around 13 meV) in SrTiO$_3$, (ii) there are significant differences in the elastic and piezoelectric properties of PbTiO$_3$ and BaTiO$_3$ (Table II).

**IV. Discussion**

A large 70-90 meV phonon band-gap seems characteristic of $ATiO_3$ perovskite quantum paraelectrics like $SrTiO_3$. In the case of the ferroelectric instability, the covalent interactions which play a predominant role[3] lead to marked differences in the phonon spectra. Ferroelectricity in perovskites arises from a competition between the Coulomb (which favors ferroelectric) and short-ranged (which favors paraelectric) interactions[3]. This competition which contributes to several reported anomalies[2-23], also leads to the spectacular signatures in the phonon spectra. The covalent Ti-O bonding (Fig. 3) found necessary for ferroelectric behavior in perovskites[3] causes the following: (i) the large 70-90 meV phonon band-gap of paraelectric $SrTiO_3$ gets filled by the covalent Ti-O vibrations in ferroelectric $PbTiO_3$ and $BaTiO_3$, and (ii) there are giant anisotropies in the phonon spectra and electromechanical response of $PbTiO_3$ (due to the combined Ti-O and Pb-O covalent bonding). The anisotropy of the phonon spectra correlates well with ferroelectric and piezoelectric strength. These results suggest that vibrational spectroscopy can aid in the search for novel materials. The distinct phonon spectra of $PbTiO_3$, $BaTiO_3$ and $SrTiO_3$ lead to important differences in their thermodynamic properties and phase diagram.

An important objective of this study was the integration of current advances in first principles computational theory with neutron experiments to provide microscopic insights that can define new strategies for the screening of novel materials. To transform the theoretical quantum mechanical techniques[24-27] developed in the last several decades into predictive design and discovery tools, important tests of the ability of modern first principles theory in reproducing various macroscopic physical properties of ferroelectrics were of interest. In this context, we have compared our calculations, with reported[15-23] Raman and infrared data (Table I), inelastic neutron data[20] (Figs. 1, 2), Brillouin data[35] (Table II) and synchrotron x-ray data[16] of three classic perovskites in addition to comparing with our experimental neutron data. Our calculations are in good agreement with these experimental studies. The theoretical studies predict giant anisotropies in the phonon dispersion relations and electromechanical response of $PbTiO_3$ which arise due to the strongly "directional" character of the covalent interactions found necessary for ferroelectricity. This anisotropy in the phonon dispersion relations obtained from first principles calculations of $PbTiO_3$ (particularly for the high energy modes which have not been earlier studied [20]) is measurable experimentally via future single crystal inelastic neutron scattering and inelastic x-ray scattering measurements. Anisotropy in the phonon spectra of $PbTiO_3$ obtained from LO-TO (longitudinal optical- transverse optical) splittings have been measured using single-crystal infrared spectroscopic studies and are found to be in good agreement with our calculations (Table I).

**V. Conclusions**

In summary, we report very successful tests of the ability of first principles theory in reproducing the experimental data of phonon dispersion relations, density of states and electromechanical response of three classic perovskites $PbTiO_3$, $BaTiO_3$ and $SrTiO_3$. We obtain an important correlation between the phonon spectra and ferroelectricity in perovskites, which can guide future efforts at custom designing still more effective piezoelectrics for applications. A large phonon band-gap seems characteristic of $ATiO_3$ perovskite quantum paraelectrics, while anisotropy of the phonon spectra correlates well with ferroelectric strength. Distinct bonding characteristics in the ferroelectric and paraelectric phases give rise to these vibrational signatures. Although, we have studied only three classic perovskites, since the vibrational signatures we obtain arise due to fundamental differences in the physics of ferroelectrics (covalently bonded) and paraelectric systems, we believe that these correlations in phonon spectra could be universal. Our results suggest that vibrational spectroscopy can aid in the search for novel materials. We hope these studies will stimulate interest in systematic investigations of other ferroelectrics to examine the universalities of our observations. The excellent agreement between theory and experiments, demonstrate the intrinsic power of first principles quantum mechanical calculations for deriving various key properties of these materials.

*Acknowledgements*
N.C. thanks R. E. Cohen and S.L. Chaplot for discussions. This research used the supercomputing resources at the Bhabha Atomic Research Centre (BARC) and the Center for Piezoelectrics by Design, College of William and Mary (E.J.W.). The work at Argonne National Laboratory was supported by the Office of Basic Energy Sciences, Division of Materials Sciences, U.S. Department of Energy, under Contract No. W-31-109-ENG-38.

**Table I**: Calculated structural parameters and long-wavelength phonon frequencies ($\omega$ cm$^{-1}$) of tetragonal PbTiO$_3$ and rhombohedral BaTiO$_3$ compared with reported experimental[1(b), 21, 22(a)] x-ray and neutron diffraction, Raman and infrared data. For tetragonal PbTiO$_3$, the $z$ atomic coordinates are given in lattice units. The $R3c$ structure of BaTiO$_3$ is defined by the lattice constant $a$ (Å), rhombohedral angle (in degree) and atomic displacements relative to ideal cubic positions $\Delta$ (Å). The calculated polyhedral (PbO$_8$, BaO$_{12}$) and TiO$_6$ octahedral distortions and bond-lengths of PbTiO$_3$ and BaTiO$_3$ are in good agreement with the experimental data. Symmetry analysis of the zone-center $\Gamma$-point phonon modes for the $P4mm$ space group of PbTiO$_3$ yields the classification $\Gamma$: 4A$_1$+5E+B$_1$. The E symmetry phonon modes are doubly degenerate. For rhombohedral BaTiO$_3$, the phonon modes can be classified as $\Gamma$ : 4A$_1$+5E+A$_2$. The longitudinal optic (LO) and transverse optic (TO) phonon frequencies of the polar modes are listed. *The non-polar E-symmetry phonon frequency in BaTiO$_3$. [a]Ref. [4], [b]Ref. [1(b)]; [bb]Ref.[21(a)], [c]Ref. [22(a)], [d]Ref.[26(b)], [e]Ref.[21(c)], [f]Ref.[21(b)].

|  |  | LDA Calculations *This work* | LDA calculations[4, 26(a)] | Experimental[1(b), 21, 22(a)] |
|---|---|---|---|---|
| PbTiO$_3$ |  |  |  |  |
|  | $a$ (Å) | 3.9048 | 3.9048 [a] | 3.9048[b], 3.902[bb] |
|  | $c/a$ | 1.063 | 1.063 [a] | 1.063[b], 1.065[bb] |
|  | $z(Ti)$ | 0.5378 | 0.549 [a] | 0.54[b], 0.5377[bb] |
|  | $z(O1, O2)$ | 0.6145 | 0.630 [a] | 0.612[b], 0.6118[bb] |
|  | $z(O3)$ | 0.1136 | 0.125 [a] | 0.112[b], 0.1117[bb] |
|  | Pb-O (Å) | 2.8016(4); 2.5243 (4) |  | 2.7980[bb] (4), 2.5319[bb] (4) |
|  | Ti-O (Å) | 1.9785 (4); 1.7606 (1); 2.3908 (1) |  | 1.9751[bb] (4), 1.7700[bb] (1), 2.3860[bb] (1) |
|  | A$_1$ (TO) (cm$^{-1}$) | 146, 355, 648 | 151[a], 355[a], 645[a] | 147[c], 359[c], 646[c] |
|  | A$_1$ (LO) (cm$^{-1}$) | 187, 438, 781 | 187[a], 449[a], 886[a] | 189[c], 465[c], 796[c] |
|  | E (TO) (cm$^{-1}$) | 90, 170, 275, 465 | 81[a], 183[a], 268[a], 464[a] | 88[c], 220[c], 289[c], 505[c] |
|  | E (LO) (cm$^{-1}$) | 117, 275, 416, 635 | 114[a], 267[a], 435[a], 625[a] | 128[c], 289[c], 436[c], 723[c] |
|  | B$_1$ (cm$^{-1}$) | 277 | 285 [a] | 289[c] |
| BaTiO$_3$ |  |  |  |  |
|  | $a$ (Å) | 4.0 | 4.0[d] | 4.0[e] |
|  | $\theta$ (degrees) | 89.90 | 89.85[d] | 89.90[e] |
|  | $\Delta_z(Ti)$ (Å) | 0.053 | 0.043 [d] | 0.052±12[e] |
|  | $\Delta_x(O)$ (Å) | -0.052 | -0.049 [d] | -0.044±8[e] |
|  | $\Delta_z(O)$ (Å) | -0.080 | -0.077 [d] | -0.072±8[e] |
|  | Ba-O (Å) | 2.8286 (6), 2.7335 (3), 2.911 (3) |  | 2.8287[f] (6), 2.7739[f] (3), 2.8981[f] (3) |
|  | Ti-O (Å) | 1.864 (3), 2.147 (3) |  | 1.8776[f] (3), 2.1351[f] (3) |
|  | A$_1$ (TO) (cm$^{-1}$) | 167, 295, 527 | 169[d], 255[d], 511[d] |  |
|  | A$_1$ (LO) (cm$^{-1}$) | 183, 462, 679 | 179[d], 460[d], 677[d] |  |
|  | E (TO) (cm$^{-1}$) | 162, 240, 468, 296* | 164[d], 206[d], 472[d], 293*[,d] |  |
|  | E (LO) (cm$^{-1}$) | 176, 440, 688, 296* | 175[d], 443[d], 687[d], 293*[,d] |  |
|  | A$_2$ (cm$^{-1}$) | 274 | 278[d] |  |

**Table II**. Calculated spontaneous polarization P (C/m$^2$), elastic constants $c_{ij}$ (GPa) and piezoelectric tensors $e_{ij}$ (C/m$^2$) of tetragonal PbTiO$_3$ (PT) and rhombohedral BaTiO$_3$ (BT) compared with reported Brillouin[35,36] and synchrotron x-ray diffraction[21(d)] data. $\varepsilon^\infty$ and $\varepsilon^0$ represent electronic and zero frequency dielectric tensors (in units of $\varepsilon_o$), respectively. For BaTiO$_3$ the tensors are in their hexagonal coordinate system, with the $z$-axis along the cubic $(111)$ direction. The spontaneous polarization for PbTiO$_3$ and BaTiO$_3$ are respectively, along the cubic [001] and [111] directions. *Ref. [21(c)].

|  | P | $c_{11}$ | $c_{33}$ | $c_{44}$ | $c_{66}$ | $c_{12}$ | $c_{13}$ | $c_{14}$ | $c_{65}$ | $e_{21}$ | $e_{31}$ | $e_{33}$ | $e_{15}$ | $e_{16}$ | $\varepsilon^\infty_{xx}$ | $\varepsilon^\infty_{zz}$ | $\varepsilon^0_{xx}$ | $\varepsilon^0_{zz}$ |
|---|---|---|---|---|---|---|---|---|---|---|---|---|---|---|---|---|---|---|
| **PT** |  |  |  |  |  |  |  |  |  |  |  |  |  |  |  |  |  |  |
| Expt.[35] | 0.75* | 237 | 60 | 69 | 104 | 90 | 70 |  |  |  | 2.1 | 5.0 | 4.4 |  |  |  | 102 | 34 |
| Calc. | 0.78 | 235 | 45 | 47 | 99 | 95 | 69 | 0 | 0 | 0 | 2.1 | 4.4 | 6.6 | 0 | 7.6 | 6.9 | 143 | 26 |
| Calc.[11] |  | 230 | 47 | 47 | 99 | 96 | 65 |  |  |  | 2.1 | 4.4 | 6.6 |  |  |  |  |  |
| **BT** |  |  |  |  |  |  |  |  |  |  |  |  |  |  |  |  |  |  |
| Calc. | 0.28 | 251 | 236 | 37 | 89 | 73 | 36 | 45 | 45 | 2.3 | 2.3 | 3.7 | 4.6 | 2.3 | 6.1 | 5.5 | 51 | 27 |
| Calc.[26(b)] |  | 277 | 264 | 48 | 99 | 79 | 41 | 45 | 45 | 2.9 | 3.0 | 4.4 | 5.5 | 2.9 | 6.2 | 5.8 | 69 | 37 |

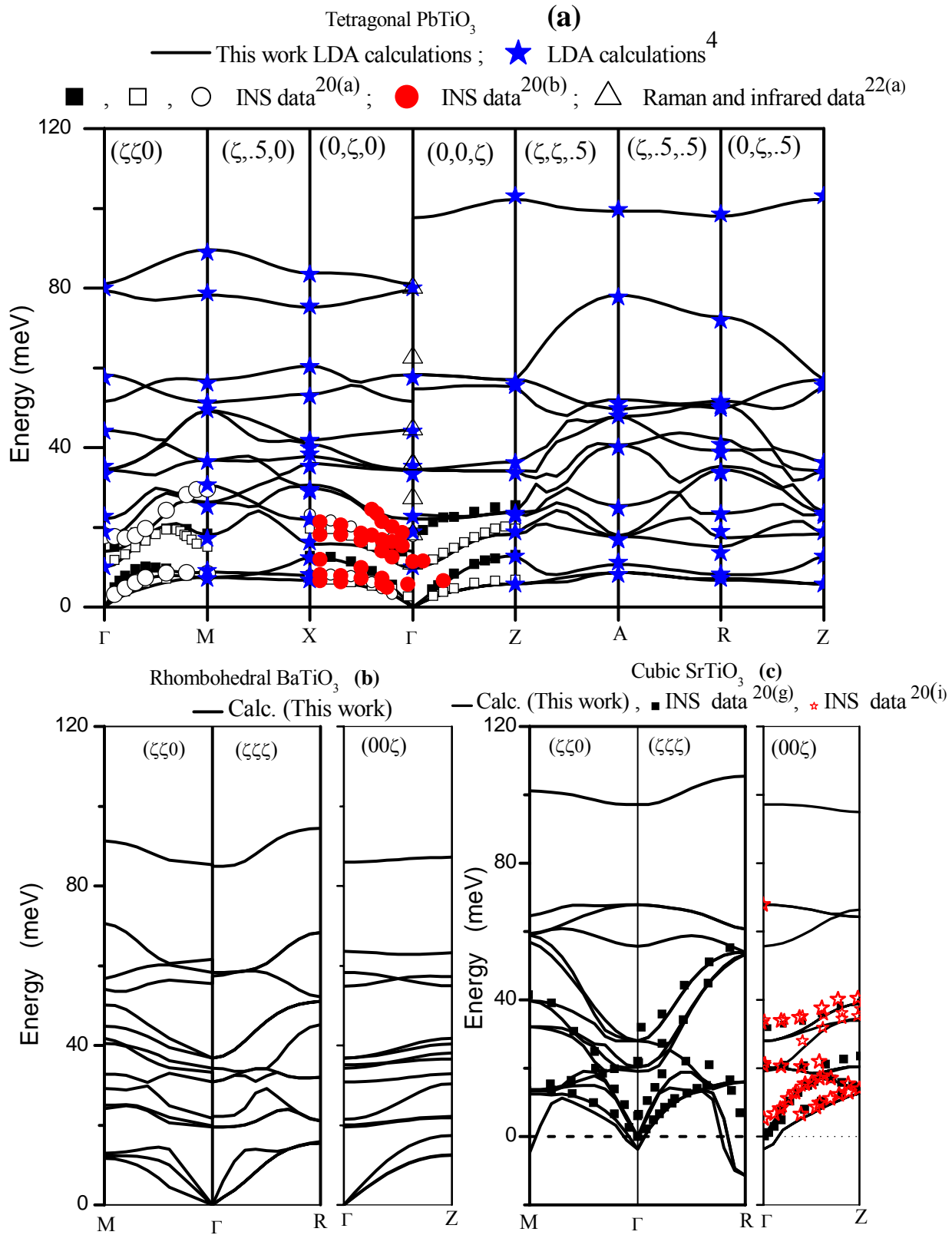

Fig. 1. (Color online). (a) Computed phonon dispersion relations (full line) of tetragonal $PbTiO_3$ (a), rhombohedral $BaTiO_3$ (b) and cubic $SrTiO_3$ (c) compared with reported experimental inelastic neutron scattering (INS) single crystal data[20], optical long wavelength data[22(a)] and reported first principles calculations[4].

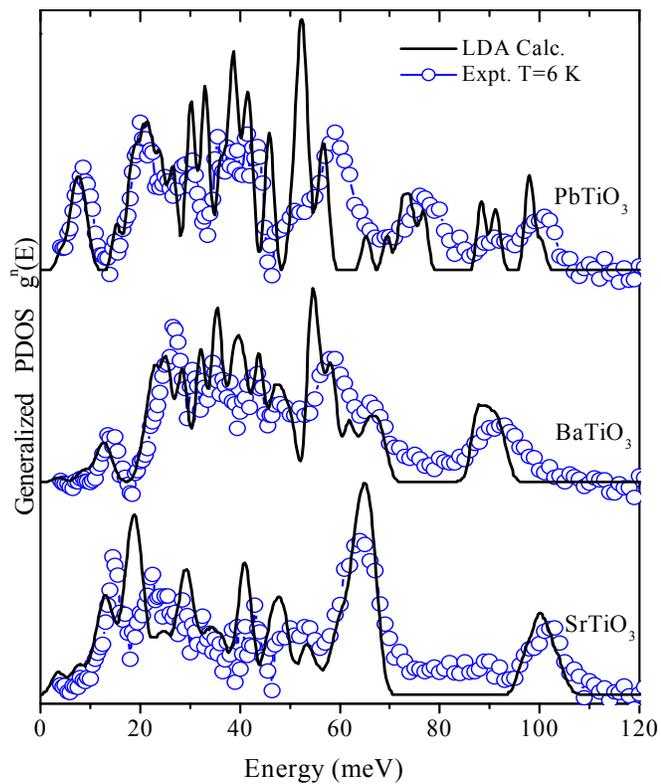

Fig. 2 (Color online). Comparison of the computed generalized phonon density of states with the measured inelastic neutron spectra (T=6 K) of tetragonal $PbTiO_3$, rhombohedral $BaTiO_3$ and the quantum paraelectric $SrTiO_3$.

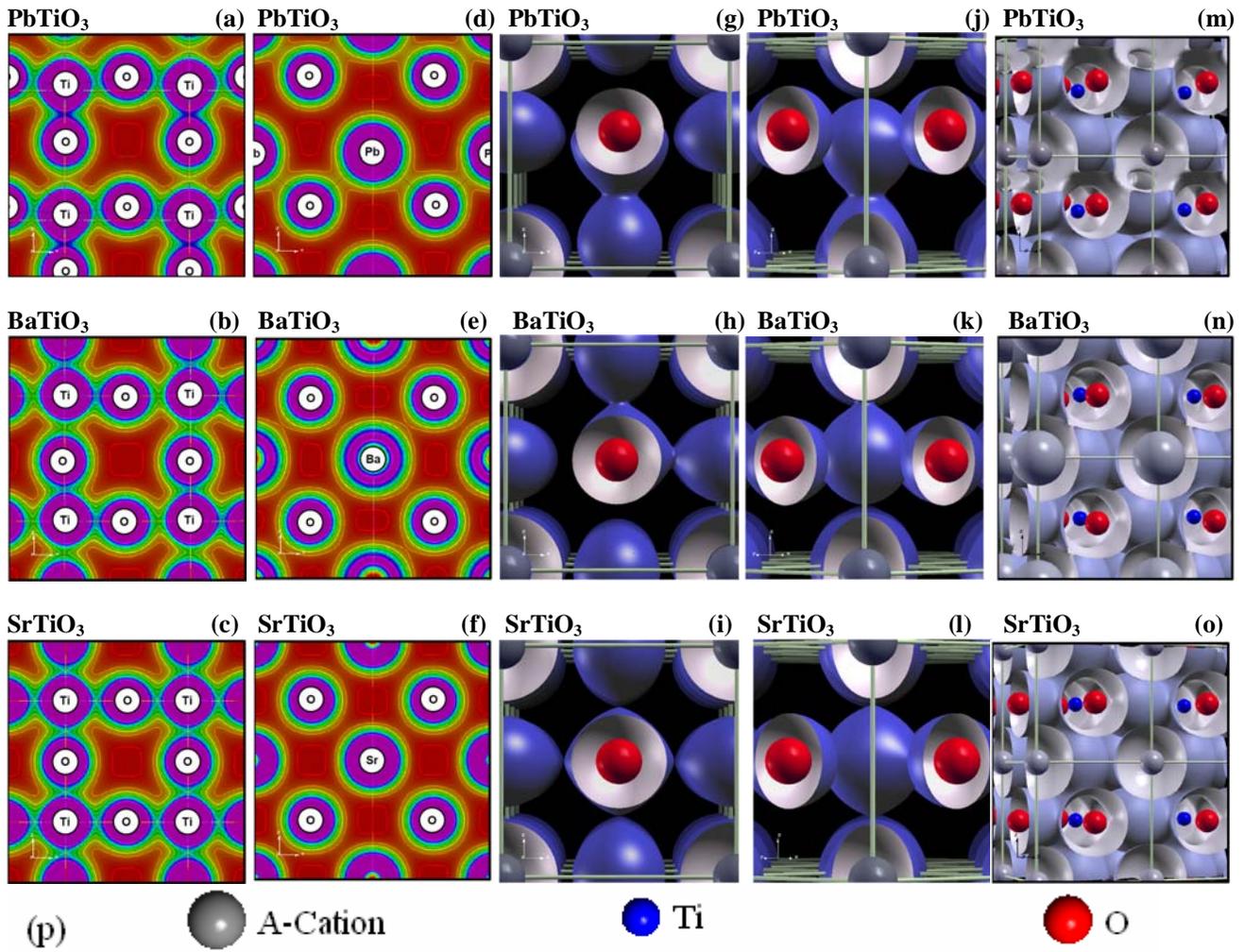

Fig. 3 (Color online). The computed charge densities (a-f) and isosurface plots (g-o) of tetragonal PT, rhombhohedral BT and cubic ST displayed using the code *xcrysden*[37]. All densities were computed from the fully relaxed zero pressure LDA structures. The charge densities in the [010] Ti-O plane (a-c) and the [010] Pb-O plane (d-f) are shown. The 0.12 a.u. isosurfaces viewed down [100] and [110] are shown in (g-i) and (j-l), respectively, while the 0.03 a.u. isosurfaces are shown in (m-o). (p) The symbols used in (g-o) above.

Fig. 4. The computed total and partial density of states of tetragonal PbTiO$_3$, rhombohedral BaTiO$_3$ and cubic SrTiO$_3$. For cubic SrTiO$_3$, effects from phonon instabilities are neglected.

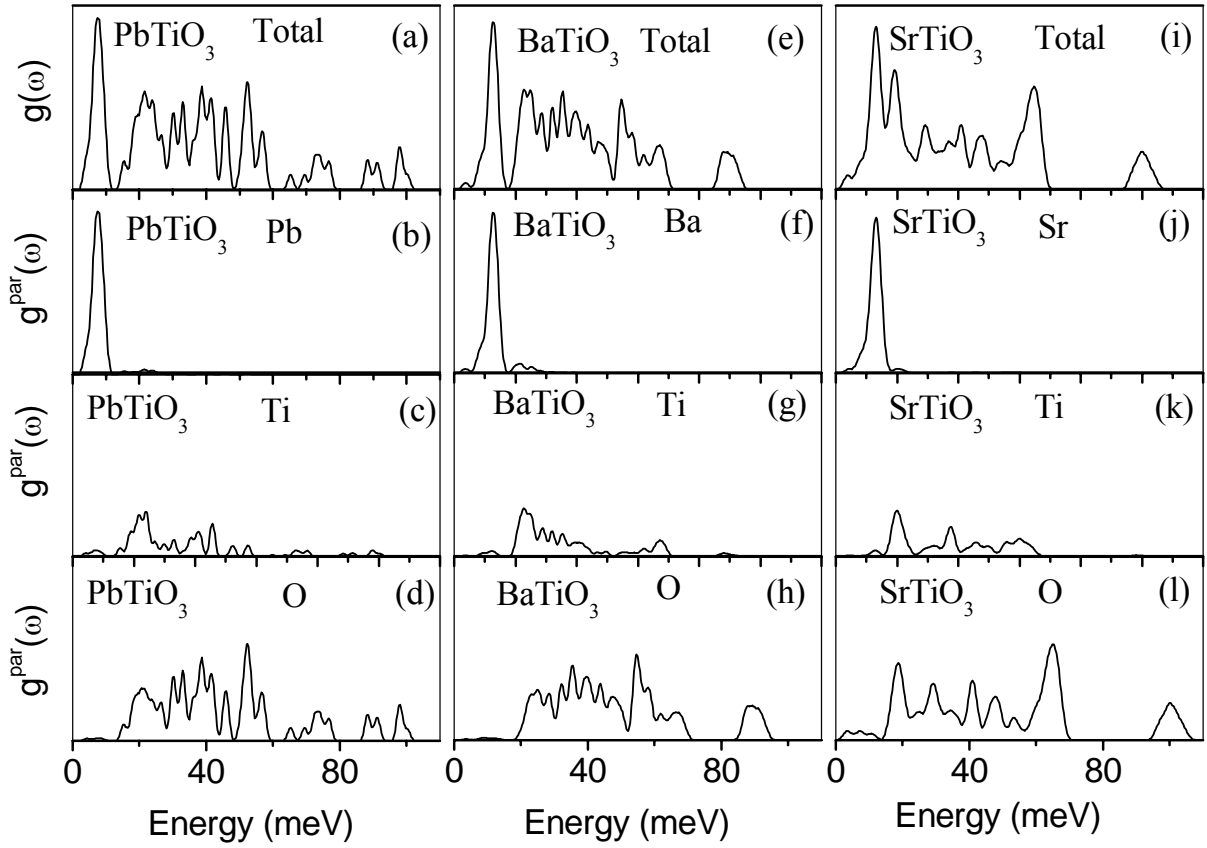